\documentclass{aastex6}
\begin{document}

\title{Korean VLBI Network Calibrator Survey (KVNCS): 1. Source catalog of KVN single dish flux density measurement in the K and Q bands }

\author{Jeong Ae Lee\altaffilmark{1,2}, Bong Won Sohn\altaffilmark{1,2,4}, Taehyun Jung\altaffilmark{1,2}, Do-Young Byun\altaffilmark{1,2} and Jee Won Lee\altaffilmark{1,3}}

\altaffiltext{1}{Korea Astronomy and Space science Institute, 776, Daedeokdae-ro, Yuseong-gu, Daejeon, Republic of Korea, 34055}
\altaffiltext{2}{Korea University of Science and Technology, 217, Gajeong-ro, Yuseong-gu, Daejeon, Republic of Korea, 34113}
\altaffiltext{3}{Kyunghee University, 1732, Deogyeong-daero, Giheung-gu, Yongin-si, Gyeonggi-do, Republic of Korea, 02447 }
\altaffiltext{4}{Corresponding author, bwsohn@kasi.re.kr}

\email{jalee@kasi.re.kr}

\begin{abstract}
We present the catalog of the KVN Calibrator Survey (KVNCS). This first part of the KVNCS is a single dish radio survey conducted at 22 (K band) and 43 GHz (Q band) simultaneously using the Korean VLBI Network (KVN) from 2009 to 2011. A total 2045 sources selected from the VLBA Calibrator Survey (VCS) with an extrapolated flux density limit of 100 mJy at K band. The KVNCS contains 1533 sources in the K band with a flux density limit of 70 mJy and 553 sources in the Q band with a flux density limit of 120 mJy; it covers the whole sky down to $-$32.$\degr$5 in declination. Five hundred thirteen sources were detected in the K and Q bands, simultaneously; $\sim$76$\%$ of them are flat-spectrum sources ($-$0.5 $\leq$ $\alpha$ $\leq$ 0.5). From the flux--flux relationship, we anticipated that the most of the radiation of many of the sources comes from the compact components. Therefore, the sources listed in the KVNCS are strong candidates for high frequency VLBI calibrators. 
\end{abstract}


\section{Introduction}
 A large proportion of compact radio sources have flat or inverted spectra \citep[e.g.,][]{Kellermann:1981ARA&A,Zensus:1997ARA&A,Gurvits:1999A&A,Chen:2009ApJ,deZotti:2010A&ARv,Massardi:2011MNRAS,Mantovani:2011A&A}. This implies that these sources are optically thick at the observed radio frequencies. Very long baseline interferometry (VLBI) calibrators are also compact sources. The majority of sources in the Very Long Baseline Array (VLBA) Calibrator Survey (VCS) have flat spectra at 2.3 (S band) and 8.4 (X band) GHz \citep{Beasley:2002ApJS, Fomalont:2003AJ, Petrov:2005AJ, Petrov:2006AJ, Kovalev:2007AJ, Petrov:2008AJ}. However, VLBI calibrators at higher frequency ($>$20 GHz) are severely rare compared to those at lower frequency. For example, 858 VLBI calibrators in the K band ($\sim$20 GHz) are known \citep{Petrov:2007AJ, Lanyi:2010AJ, Petrov:2011AJ, Petrov:2012AJ, Petrov:2012MNRAS}, while $\sim$3800 sources are listed in the VCS in the S and X bands. High$-$frequency VLBI observations are useful to understand the physical processes in the vicinity of supermassive black holes of active galactic nuclei (AGN), because a study of optically thin region from the synchrotron radiation of compact radio sources is possible. In addition, radio sources become more compact in structure so that astrometric errors in the celestial reference frame (CRF) would be minimized \citep[e.g.][]{Fey:1997ApJS, Ma:1998AJ, Fey:2004AJ, Lanyi:2010AJ, Charlot:2010AJ}. Together with the successful operation of the Gaia spacecraft during the first two years of sky survey, high$-$frequency VLBI sources matching with Gaia samples will be important to compare the radio and optical reference frames \citep[e.g.][]{Bourda:2010A&A, Bourda:2011A&A, Jacobs:2014ivs}. 
 
There is an extensive blind survey of the southern sky that identified 5808 sources at 20 GHz (AT20G) in order to support subtraction of foreground objects for the measurement of the Cosmic Microwave Background \citep{Murphy:2010MNRAS}. In the northern sky, \cite{Righini:2012MNRAS} conducted the K-band Northern Wide Survey and identified 73 sources at declination $>$ 72.$\degr$3 ($\sim$880 degree$^{2}$). Recently, \cite{Jacobs:2014ivs} constructed a new compact radio survey at X/Ka (8.4/32 GHz) bands in order to improve the accuracy of celestial reference frame and detected 631 sources. However, the sky coverage of known K band calibrators is about 49$\%$ of the whole sky by assuming the circles of 5$\degr$ radius as shown in Figure 1, while the VCS sources in the S and X bands cover the full sky above $-$40$\degr$ in declination with this radius \citep{Petrov:2008AJ}. The critical separation angle, 5$\degr$ was determined, referring to \cite{Dodson:2014AJ}. They showed a feasibility of source frequency phase referencing (SFPR) in KVN, considering of 5$\degr$.9 separation angle between a target and a calibrator in Korean VLBI Network (KVN). We therefore performed 21.7 (K band) and 42.4 (Q band) GHz single dish survey observations in order to provide a wider sky coverage of possible VLBI target and calibrator sources at those frequencies (or even higher frequencies), particularly for the KVN.

 The KVN is composed of three 21-m telescopes in Seoul, Ulsan and Jeju, Korea. Its notable characteristic is the multifrequency receiver system, which makes it possible to observe the radio sources at four different frequencies [22 (K), 43 (Q), 86 (W), and 129 (D bands) GHz] simultaneously \citep[e.g.,][]{Han:2008IJIMW, sslee:2011PASP, sslee:2014AJ}. In particular, this system is very efficient for high frequency VLBI observations when the frequency phase transfer (FPT) method is applied to compensate for the atmospheric coherence loss; this method uses lower frequency phase solutions (e.g., in the K band) to compensate for higher frequency ones (e.g., in the Q/W/D bands), because of the non-dispersive nature of the atmosphere with regard to radio systems \citep[e.g.][]{Jung:2011PASJ, Rioja:2014AJ, Rioja:2015AJ}. As a result, noticeable improvements of coherence time and signal-to-noise ratio (SNR) at high frequency VLBI observations have been demonstrated \citep[e.g.][]{Jung:2012evn, Jung:2014evn}.
 
 For the practical use of the FPT for higher frequencies (Q/W/D-band) VLBI observations with the KVN, increasing available sources in the K and Q bands is very important in the northern sky because the detection of target or calibrator sources at the lower frequencies (e.g. in the K or Q bands) as a reference for the atmospheric calibration is essential. The increase of the detection at high frequencies by the FPT will be able to extend our understanding of radio sources from centimeter to millimeter wavelengths in VLBI; for example, a statistical study of a spectral energy distribution (SED) of AGNs based on the simultaneously measured flux densities at a range from 20 to 130 GHz, and high frequency astrometric applications \cite[e.g.][]{Rioja:2015AJ}. We therefore performed the KVN Calibrator Survey (KVNCS), which aims to measure the single dish flux densities of $\sim$2500 selected sources in the K and Q bands simultaneously in order to utilize as VLBI calibrators of KVN. 

 In the next section, we explain the selection criteria for the $\sim$2500 sources. The observations and data reductions are described in section 3. In section 4, the observational results and their analysis are presented. Finally, we summarize our findings and future prospects in section 5. The full catalog of the KVNCS is also provided in the appendix.

\section{Source selection}
A total of 2503 sources with an extrapolated total flux density (hereafter, S$_{K}^{S-X}$) greater than 100 mJy in the K band were selected from VCS1 to VCS5 \citep{Beasley:2002ApJS, Fomalont:2003AJ, Petrov:2005AJ, Petrov:2006AJ, Kovalev:2007AJ}. As the most widely used catalog of VLBI calibrators, the VCS contains $\sim$3800 radio sources that show mostly a compact structure and flat spectra sources with greater than $-$40$\degr$ in declination. S$_{K}^{S-X}$ was calculated, assuming a power-law spectrum from the total flux densities in the S and X bands \citep{Sohn:2009}. The flux density limit of 100 mJy was a 
selection criterion, which gives $>$8$\sigma$ at a baseline sensitivity of the KVN at K band\footnote{KVN status report: http://kvn.kasi.re.kr/status$\_$report/}. A declination limit of $-$32.$\degr$5 was determined so that all sources are observed at more than 20 $\degr$ above the horizon at transit. For the first observation as a pilot observation, 595 relatively bright sources that have S$_{K}^{S-X}$ higher than 500 mJy in the K band were selected, and 493 sources ($\sim$83$\%$) were successfully detected with the first single dish observations from KVNCS1.0.1 to KVNCS1.1.3 in the K and Q bands. Then, we performed a single dish observation toward 1450 of the remaining 1908 sources, the flux density of which ranges between 100 mJy and 1 Jy. In order to avoid concentrated on the specific regions, the 1450 sources were selected by considering the sky coverage of the detected source density, which was calculated by the Delaunay triangulation method\footnote{This method was developed by Boris Delaunay in 1934 and is well known in mathematics and computational geometry. The triangles in Delaunay triangulation method minimizes the summation of the interior angles of the triangles. In addition, these triangles do not include any points except the vertices within their circumcircle (from Wikipedia). }, whenever each observation was complete. We assumed that the calibrators are located on the vertex of the triangle in the triangulation method. The target is located in the center of the circumcircle of this triangle, and the radius of this circle become the maximum separation angle between the target and the calibrators.

\section{Observation and data reduction}

 \subsection{Observation}
 We selected 2503 sources based on lower frequency VLBI flux density from VCS. A flux densities of 2045 of 2503 sources were measured by single dish observations in the K and Q bands simultaneously using the KVN Yonsei and Ulsan radio telescopes from December 2009 to March 2011. The observational information is summarized in Table 1. The cross-scan mode (cs-mode), a one-dimensional on-the-fly method, was used to obtain an accurate flux density measurement. A single cs-mode scan consists of two scans in azimuth (Az) and two in elevation (El). Each scan includes both forward and backward scans. We selected the parameters for the cs-mode considering the telescope beam size, time scales of the sky power variation, and a hardware and software limitations of the KVN system. The applied scan speed, the data sampling interval and the scan length are mainly 65$\arcsec$/sec, 0.1 ms, and 13$\arcmin$, respectively, which yield 4 (K band) and 2 (Q band) seconds full$-$beam cross time. We can remove sky power variation of which time scale is longer than the full$-$beam cross time by fitting off$-$source data. Assuming a Gaussian beam pattern, the on-source integration time of a cross scan are 8 and  4 seconds in the K and Q bands, respectively. The total on-source time for each source is considered based on the estimated source flux density and is calculated by multiplying by the number of scans. The main-beam sizes of the KVN telescopes are around 126$\arcsec$ and 63$\arcsec$ in the K and Q bands, respectively \footnote{KVN status report: http://kvn.kasi.re.kr/status$\_$report/}. The mean beam sizes in the K and Q bands are quite consistent with the known ones. However, the estimated beam sizes and pointing offsets have large errors for faint sources and under the bad weather conditions or rapid sky variation \citep[e.g.,][]{Fante:1975IEEE}. We used them to eliminate low-quality data with 30$\%$ differences in beam size from the known ones and with the pointing offsets larger than 25$\arcsec$ arbitrarily. The standard deviations in the beam size and pointing offset data are 7$\%$ (K) and 20$\%$ (Q) and 4 $\arcsec$ (K) and 4 $\arcsec$ (Q) in Az and 10$\%$ (K) and 19$\%$ (Q) and 6.4 $\arcsec$ (K) and 6.0 $\arcsec$ (Q) in El, respectively. The observing frequencies were 21.7 GHz for the K band and 42.4 GHz for the Q band with 512 MHz bandwidth. A measured root-mean-square (RMS) error is an order of magnitude higher than estimated thermal one. This is usual in practice.
 Hot/cold load calibration was performed in order to determine the antenna temperature scale from KVNCS1.0.1 to KVNCS1.2.1. For hot/cold load calibration, we used two microwave absorbers, one in room temperature of $\sim$292K and the other immersed in liquid nitrogen of 80K, as hot and cold loads, respectively. From KVNCS1.2.2, we used chopper$-$wheel method which uses a sky as a cold load together with a microwave absorber in room temperature \citep{Kutner:1981ApJ,Mangum:2000Manual}. The results from these two calibration methods were consistent each other within $\sim$2$\%$ uncertainty. The system temperature and zenith optical depth of each observation are presented in Table 1. The sky opacity was corrected using the sky-dipping method based on observations every hour.  3C 286 was observed as a flux calibrator every 1.5 h to convert the measured antenna temperature into the flux. The flux densities of 3C 286 are 2.64 and 1.51 Jy in the K and Q bands, respectively. These were measured with a brightness model of Mars at the KVN Yonsei observatory (Sohn, in prep.).

 \subsection{Data reduction}
We developed an analysis program for the KVN single dish flux density measurements and the following procedures were applied: (a) Extraction of bad scans due to the weather conditions or instrumental spuriousness, (b) linear baseline fitting to estimate the RMS noise level and to eliminate sky level, (c) Gaussian fitting to measure the antenna temperature and to correct pointing offsets deviating from the Gaussian fitted center position using equations (1) and (2) \citep{Fuhrmann:diss2004}. 
\begin{equation}
\centering
(T^{*}_{a,Az})^{\prime} = T^{*}_{a,Az} \cdot \exp[4\ln 2 \frac{x^{2}_{El}} {\theta^{2}_{El}}] 
\end{equation}
\begin{equation}
\centering
(T^{*}_{a,El})^{\prime} = T^{*}_{a,El} \cdot \exp[4\ln 2 \frac{x^{2}_{Az}}{\theta^{2}_{Az}}]
\end{equation}
$T^{*}_{a,Az}$ and $T^{*}_{a,El}$ are the measured antenna temperature corrected for atmospheric attenuation along the Az and El axes in the cs-mode, respectively. Further, $x_{Az}$ and $x_{El}$ are pointing offsets, and $\theta_{Az}$ and $\theta_{El}$ are the half-power beam width in arcseconds. After the pointing correction, $(T^{*}_{a,Az})^{\prime}$ and $(T^{*}_{a,El})^{\prime}$ were averaged. (d) The error propagation of the antenna temperature was calculated using equation (3).
\begin{equation}
\centering
\sigma_{(T^{*}_{a})^{\prime}} = (T^{*}_{a})^{\prime} \sqrt{[\frac{\sigma^2_{\Delta G}}{{\Delta G}^2} + \frac{\sigma^2_{T^{*}_{a}}}{(T^{*}_{a})^2}]}
\end{equation}
$(T^{*}_{a})^{\prime}$ is the corrected and averaged $T^{*}_{a}$ for the pointing offset and in Az and El. $\sigma_{T^{*}_{a}}$ represents the uncertainty of $T^{*}_{a}$. $\Delta$G is the temporal variation of the antenna gain, and its uncertainty is written as $\sigma_{\Delta{G}}$. The gain curves of the KVN radio telescopes are very flat for elevations of 20 to 80$\degr$ in the K and Q bands. The gain variation along this elevations were 2.5$\%$ (K) and 2.0$\%$ (Q) at KYS and 1.5$\%$ (K) and 4.5$\%$ (Q) at KUS $^{3}$. Thus, the elevation dependence of the gain variation was ignored. $\Delta$G was obtained by using the ratio of the mean $(T^{*}_{a})^{\prime}$ to $(T^{*}_{a})^{\prime}$ of 3C 286. However, the error was dominated by the thermal random noise. The flux density conversion factors were determined from the ratio of the flux densities to $(T^{*}_{a})^{\prime}$ of 3C 286 in the  K and Q bands (see Table 1), which were obtained using the NRAO Mars emission model \citep{Butler:2001Icar}. To check the results, we compared these conversion factors with those obtained for planets \citep{sslee:2011PASP}, which have $\sim$10$\%$ uncertainty, and found that they are consistent within 10$\%$ and 8$\%$ in the K and Q bands, respectively. Finally, the conversion factors obtained for 3C 286 were applied to estimate the source flux densities.

\section{Results and Discussion}
 \subsection{Flux density measurement}
 Among 2043 sources, the flux densities of 1533 (75$\%$) and 533 (27$\%$) sources with 3$\sigma$ noise levels of 66 (K band) and 108 (Q band) mJy were successfully measured. Their median flux densities were 397 and 588 mJy and the lowest flux densities were $\sim$70 and $\sim$120 mJy in the K and Q bands, respectively. Among them, the flux densities of 513 sources were measured in the K and Q bands simultaneously. The distribution of the measured flux densities are shown in Figure 2 and the measured flux densities are listed in Table 2.

 The luminosity distribution of 1138 sources as a function of the redshift ($z$) are plotted in Figure 3. Archival redshifts were taken from NASA/IPAC Extragalactic Database (NED) and the Sloan Digital Sky Survey (SDSS) DR13. Only 23 sources were given as the photometric redshifts and their mean uncertainty are about 0.618. These luminosities are calculated with $H_{0}$ = 73 km $s^{-1}$ $Mpc^{-1}$ and $\Omega_{m}$ = 0.27 at 21.7 GHz \footnote{http://ned.ipac.caltech.edu/}, according to the arbitrarily spectral indices ($\alpha$ = $-$1.0 and 0.0, $S$ $\thicksim$ $\nu^{\alpha}$, where $S$ is the flux, and $\nu$ is the observed frequency). We denote spectra as steep ($\alpha$ $<$ $-$0.5), flat ($-$0.5 $\leq$ $\alpha$ $\leq$ 0.5) and inverted ($\alpha$ $>$ 0.5) in this study. This figure reflects that our data are those of flux-limited samples, and they show the distribution of high-power radio sources, which ranges from $10^{24}$ to $10^{29}$ W $Hz^{-1}$ at 21.7 GHz (the observing frequency). According to the number distribution with respect to $z$ in the bottom right panel in Figure 3, the peak is located at the $z$ $\sim$ 1. That of bright quasars is known to be around $z$ $\sim$ 1 \citep{White:2000ApJS}. In addition, these distributions in four-$z$ bins are shown in Figure 4. Gray-filled and black-hatched bars indicate each distribution according to $\alpha$ = $-$1.0 and 0.0, respectively. The distributions according to $\alpha$ are similar in the low-$z$ region ($z$ $<$ 0.5), whereas they differ in the high-$z$ region ($z$ $\geq$ 0.5).

 \subsection{Comparison with the VLBI flux densities in the S, X, K, and Q bands}
We compared the VCS flux densities in the S and X bands with those of the KVNCS in the K and Q bands, because large difference, if ever, between the observed flux densities in the K and Q bands and the extrapolated flux densities in the S and X bands would mean VLBI missing flux problem, high source variability or source evolution (e.g., opacity changes). Their correlations with the weighted linear fit lines are shown in Figure 5. The linear correlation coefficients are 0.77 (S--K), 0.87 (X--K), 0.81 (S--Q), and 0.85 (X--Q). The linear fit lines are obtained using the data having a K-band flux density greater than 397 mJy (median flux) and less than 1Jy (arbitrarily) and a Q-band flux density greater than 587 mJy (median flux), because the data contain faint sources that were not detected at high frequency (e.g., the K or Q bands) but were detected at low frequency (e.g., the S or X bands). This causes a selection effect for non-detected sources (the red and black arrows), which are faint sources with a flux density less than 3$\sigma$. Thus, the sources less than the median flux density were excepted for fitting. In addition, the bright sources ($>$ 1Jy) in the K band excluded for fitting because they can always be detected, although they are highly variable. The slopes of the weighted linear fit lines are 1.04 $\pm$ 0.024 and 0.79 $\pm$ 0.005 for the S--K and S--Q bands, respectively, whereas they are 0.63 $\pm$ 0.024 (X--K) and 0.59 $\pm$ 0.005 (X--Q). The differences between the slopes in the S and X bands imply that there are missing flux densities in the X band. The flux densities measured from the VCS were calculated from the CLEAN components from VLBI observations and the flux densities for some sources were obtained within 7 (S band) and 25 (X band) M$\lambda$ in projected uv-distance \citep{Petrov:2006AJ, Kovalev:2007AJ}. Therefore, there is missing flux density in the X bands, depending on the structure of each source.

We also compared the flux densities of  232 (K band) and 76 (Q band) common sources from the KVNCS and VLBI imaging survey for the International Celestial Reference Frame at 24 and 43 GHz \citep{Charlot:2010AJ}. Figure 6 shows their flux--flux relationships in the K and Q bands. On the basis of the structure index (SI) from \cite{Charlot:2010AJ}, 195 (84$\%$) and 63 (83$\%$) sources were identified as SI $<$ 3 (compact sources), and the remaining 37 (16$\%$) and 13 (17$\%$) sources showed SI $\geq$ 3 (sources with marginally compact or extended structure) in the K and Q bands, respectively. To obtain the weighted linear fit results, the data were selected in the same way as those in Figure 5. The linear fit at SI $<$ 3 was performed for source flux densities greater than 397 mJy and less than 1 Jy. However, some compact sources show the big differences of flux density between VLBI and single dish. For instance, a source which has $\sim$4 Jy of VLBI and $\sim$1 Jy of single dish flux densities has been known as a variable, J0238+1636 which had big flux variations of around 5 Jy from 2007 to 2010 \footnote{F-GAMMA project: http://www3.mpifr-bonn.mpg.de/div/vlbi/fgamma/fgamma.html}. In addition, a source which has $\sim$0.3 Jy (VLBI) and $\sim$3 Jy (single dish) is J1849+6705. Also this has been known as a variable which shows year-scale variations $^{5}$. Therefore, these variables were not considered to fit in the K band. At SI $\geq$ 3, on the other hand, the linear fit results were calculated with all sources because the number of sources was small to fit. The slope for the compact sources is 1.07 $\pm$ 0.065, while that of extended sources is 0.66 $\pm$ 0.006 in the K band. We infer that most of the radiation from the compact sources (SI $<$ 3) comes from the compact core region, whereas the extended sources (SI $\geq$ 3) have missing flux. In the Q band, because there are few common sources, the weighted linear fit was applied to all 76 sources. We expect that most of the radiation ($\sim$80$\%$) comes from the compact component, although about 20$\%$ of the flux density in the Q band is missing.
  
 \subsection{Spectral index distributions}
 Table 3 shows the statistics of $\alpha$ for 1533 sources in the K band and 553 sources in the Q band, and 513 sources detected in the K and Q bands simultaneously. As expected, many of our sources have flat spectra in both the S and X and the K and Q bands. The distributions of the spectral indices of the S and X ($\alpha_{SX}$), X and K ($\alpha_{XK}$), X and Q ($\alpha_{XQ}$), and K and Q ($\alpha_{KQ}$) bands are shown in Figure 7. The top panel shows the distributions of $\alpha_{SX}$ and $\alpha_{XK}$ for 1533 sources and $\alpha_{XQ}$ for 553 sources. The distribution of $\alpha_{XK}$ become broader than that of $\alpha_{SX}$, but $\sim$88$\%$ of the $\alpha_{SX}$ values and $\sim$70$\%$ of the $\alpha_{XK}$ values belong to the flat spectrum ranges. In addition, the distribution of $\alpha_{XQ}$ is similar, and we expect that these are relatively bright sources. In the bottom panel, the distribution of $\alpha_{XK}$ is broader than the other distributions and shifts toward the inverted spectrum region because these sources are corrupted by flux density variability. There is an observational epoch gap between the VCS and KVNCS. However, $\alpha_{XQ}$ shows a distribution similar to that of $\alpha_{SX}$. These sources are sufficiently bright. In addition, that of $\alpha_{KQ}$ is steeper ($\sim$14$\%$) than those of $\alpha_{SX}$ ($\sim$6$\%$), $\alpha_{XK}$ ($\sim$6$\%$), and $\alpha_{XQ}$ ($\sim$3$\%$). We infer that the sources become relatively optically thin in the K and Q bands. Nevertheless, 76$\%$ of the sources show flat spectra in the K and Q bands. 

 \subsection{Sky coverage}
 Figure 8 clearly shows that our sample is fairly evenly distributed on the sky (this figure is the same as Figure 1 in \cite{jalee:2012PoS}). Assuming a spatial coherence scale of 5$\degr$, which is the separation angle between a target and the calibrators, 99$\%$ of the sky is covered by these sources above $-$32.$\degr$5 in declination (Figure 9). The sky coverage is improved by more than 20$\%$ compared to that of the existing calibrators shown in Figure 1. In the same way, if the spatial coherence scales are assumed as 2.2 and 3$\degr$, the sky coverages are improved about 28 and 38$\%$, respectively. 2.2$\degr$ means a maximum separation of VERA dual-beam \footnote{http://veraserver.mtk.nao.ac.jp/system/dualbeam-e.html} which is able to observe a target and a calibrator simultaneously and 3$\degr$ is a typical coherence scale in the K band. Hence, it is expected that there will have the possibility of improving the sky coverage of the K band calibrators with a uniform distribution, comparing to the existing one (Figure 1).

\section{Summary}
We conducted an extensive single dish survey (the KVNCS) of 2043 extragalactic radio sources in the K and Q bands using the KVN and successfully detected 1533 (75$\%$) of the sources in the K band and 553 (27$\%$) in the Q band. In addition, the flux densities of 513 sources were measured in the K and Q bands, simultaneously. This catalog is an important database for high frequency VLBI observations with the KVN and other available radio telescopes worldwide. In addition, these sources become VLBI calibrator candidates. The sky density distribution of the 1533 sources covered about 99$\%$ of the sky observable by the KVN. About 76$\%$ of the 513 sources still showed flat spectra in the K and Q bands. On the basis of the flux--flux relationship between the single dish survey and VLBI observations, we inferred that most of the radiation of many of the sources comes from the compact components. To confirm the feasibility of using these sources as reliable VLBI phase calibrators, however, VLBI fringe and imaging surveys should be performed. These VLBI follow-ups are ongoing with the KVN and  KaVA (KVN and VERA Array) in the K band. In addition, a simultaneous multiwavelength Active Galactic Nuclei survey is in progress.

\acknowledgments
 The authors appreciate the support of DRC program of Korea Research Council of Fundamental Science and Technology (FY 2013). 
We are grateful to all staff members at the KVN who helped to operate the array and to correlate the data. The KVN is a facility operated by KASI (Korea Astronomy and Space Science Institute). KVN operations are supported by KREONET (Korea Research Environment Open NETwork) which is managed and operated by KISTI (Korea Institute of Science and Technology Information). The authors are grateful for the valuable comments of the anonymous referee.

\bibliography{ref}
\clearpage

\begin{figure}
\centering
\plotone{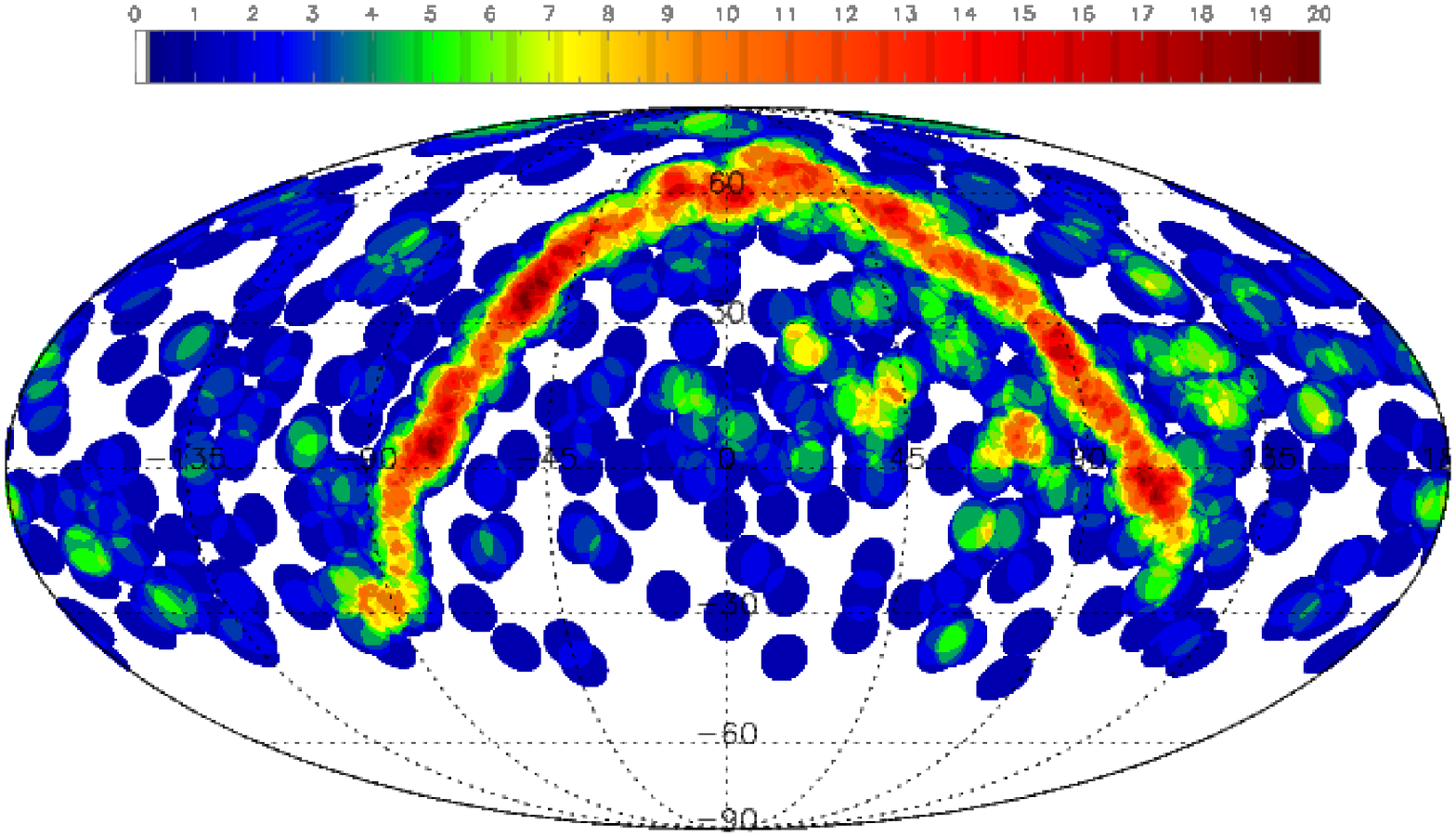}\\ 
\caption{Spatial distribution of 858 sources detected in the K band from \cite{Petrov:2007AJ}, \cite{Lanyi:2010AJ}, \cite{Petrov:2011AJ}, \cite{Petrov:2012AJ} and \cite{Petrov:2012MNRAS}. The radius of each circle is 5$\degr$, which is the separation angle between sources. The colored legend shows the values represented by overlapping circles. This map is drawn using a Mollwide equal-area projection (abscissa: right ascension [$\degr$], ordinate: declination [$\degr$]).}
\end{figure}

\begin{figure}
\centering
\plotone{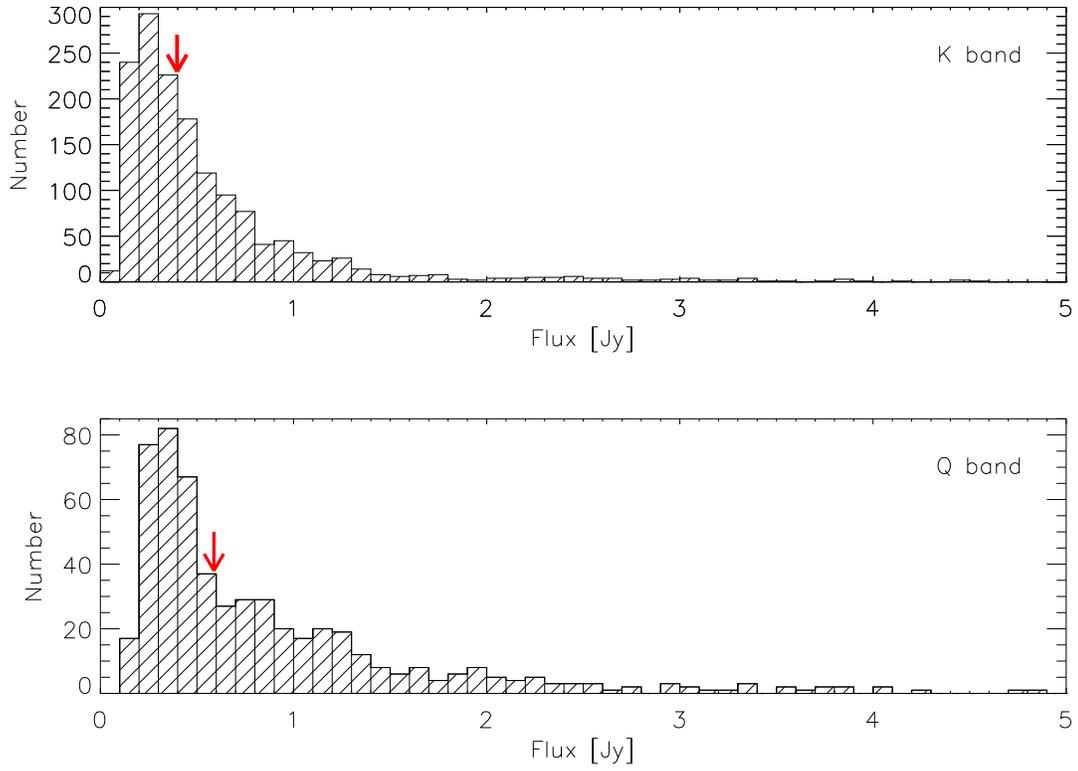}\\
\caption{Flux density distributions of measured sources ($\leq$ 5 Jy) in the K (top) and Q (bottom) bands. Red arrows indicate the median flux densities, 397 and 587 mJy in the K and Q bands, respectively, whereas the mean flux densities are 707 and 1,103 mJy, respectively. }
\end{figure}

\begin{figure} 
\centering
\plotone{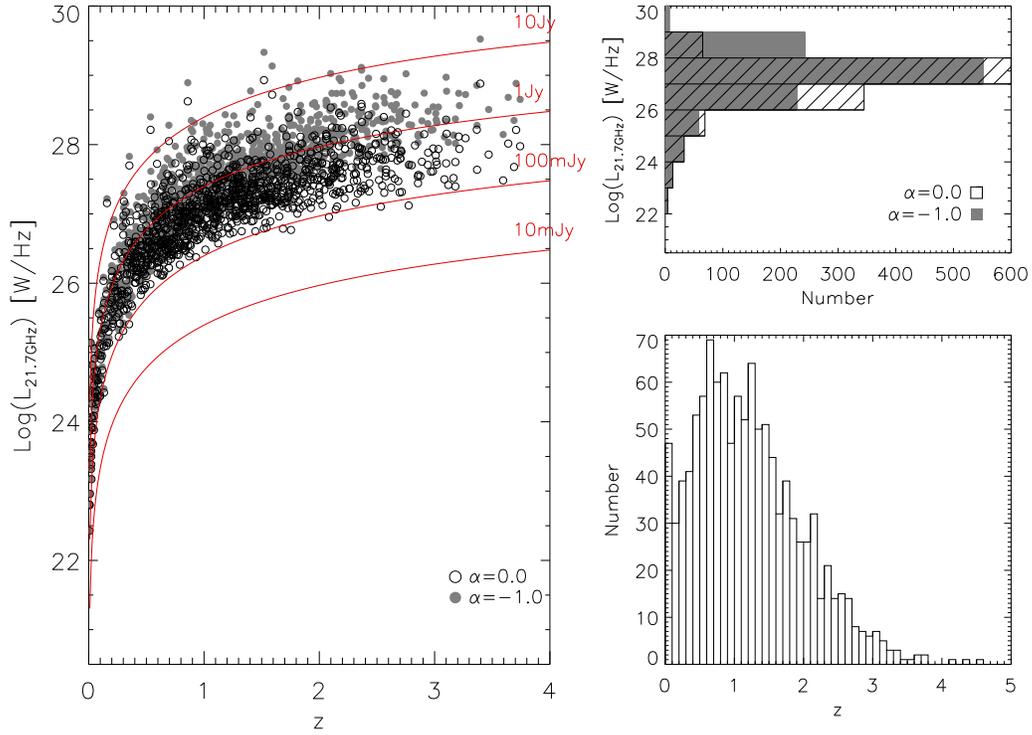}\\
\caption{Luminosity distribution of 1137 sources as a function of redshift ($z$). The $z$ values were taken from the NASA/IPAC Extragalactic Database (NED) and the Sloan Digital Sky Survey (SDSS) DR13. These luminosities are calculated with $H_{0}$ = 73 km $s^{-1}$ $Mpc^{-1}$ and $\Omega_{m}$ = 0.27 at 21.7 GHz, which is the observed frequency, according to the spectral index, $\alpha$ ($-$1.0: gray filled circles, 0: black circles). The red lines indicates the luminosity distribution at $\alpha$ = 0 as the criteria for flux densities. Top right: The number of sources according to their luminosities at $\alpha$ = $-$1 and 0. Bottom right: The number of sources according to $z$.}
\end{figure}

\begin{figure} 
\centering
\plotone{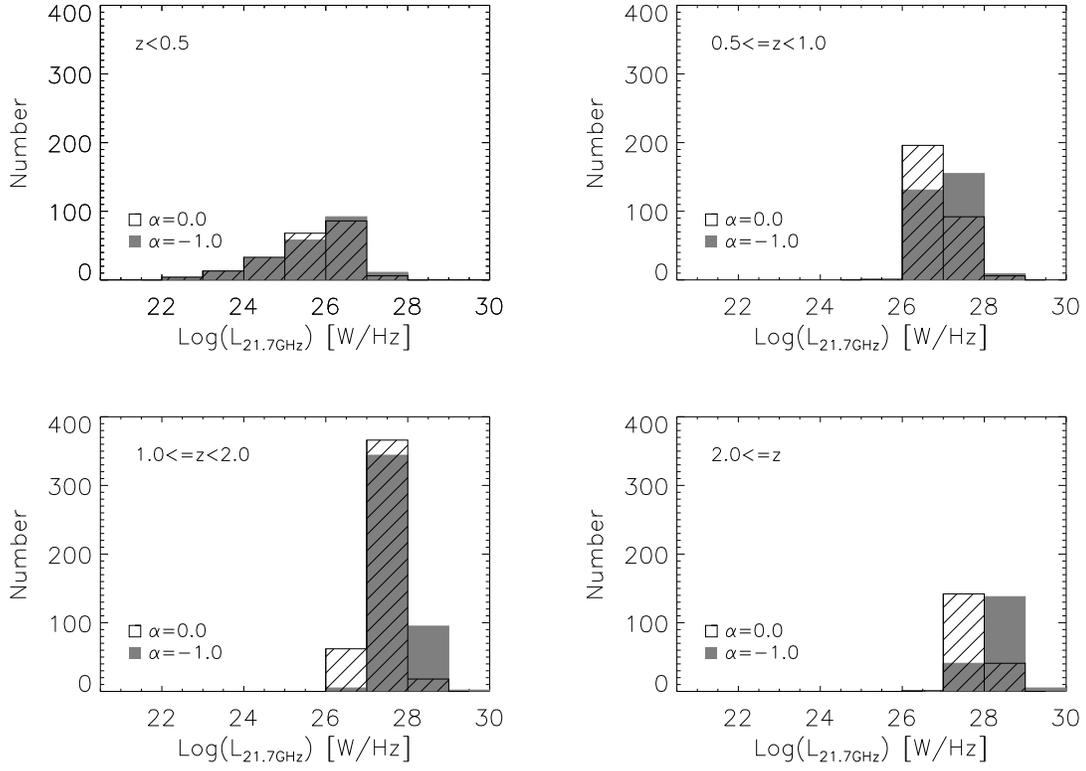}\\
\caption{Luminosity distributions in four-redshift ($z$) bins. Gray filled and black hatched bars stand for the distributions for $\alpha$ = $-$1.0 and 0.0, respectively.}
\end{figure}

\begin{figure} 
\centering
\plotone{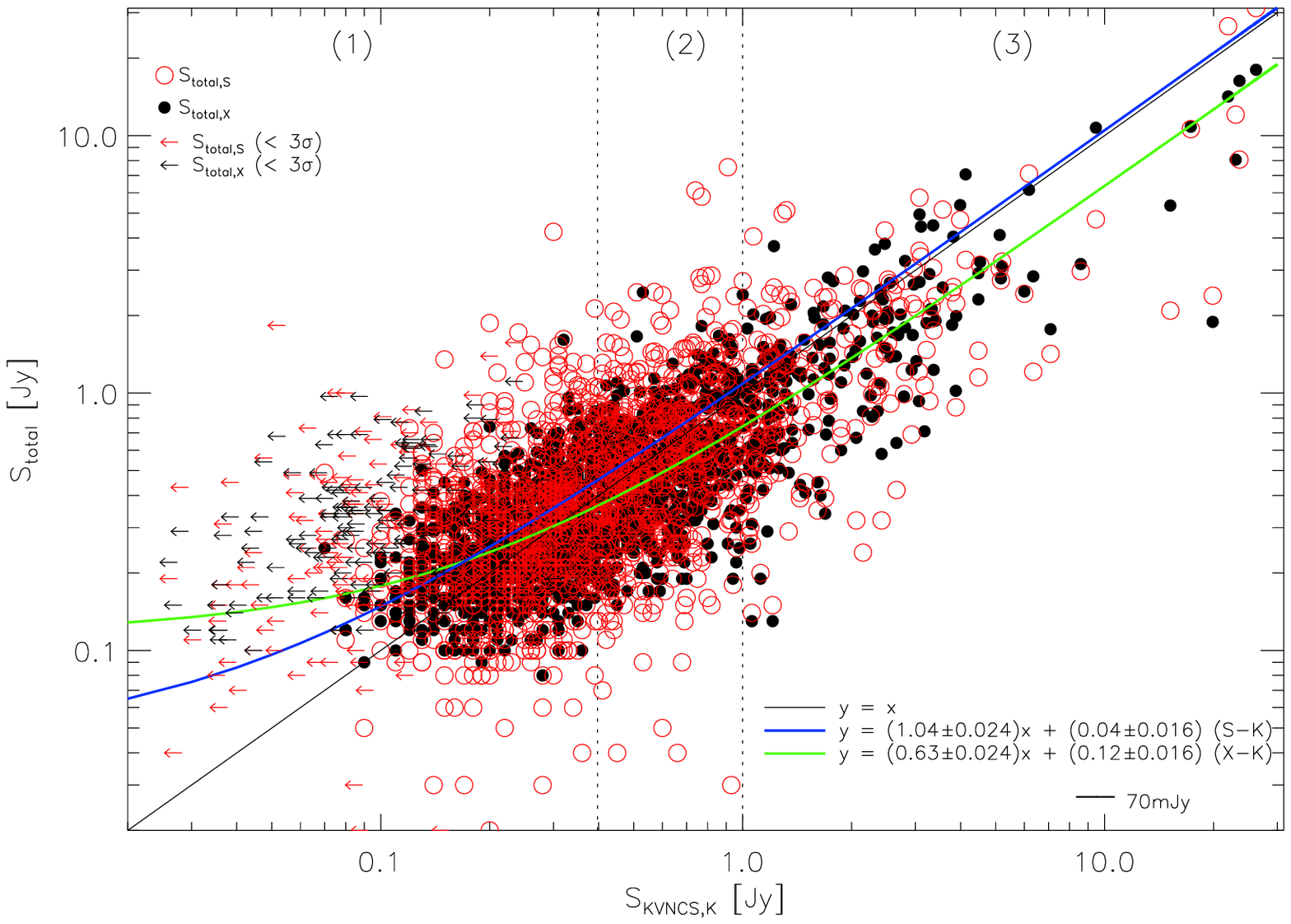}
\plotone{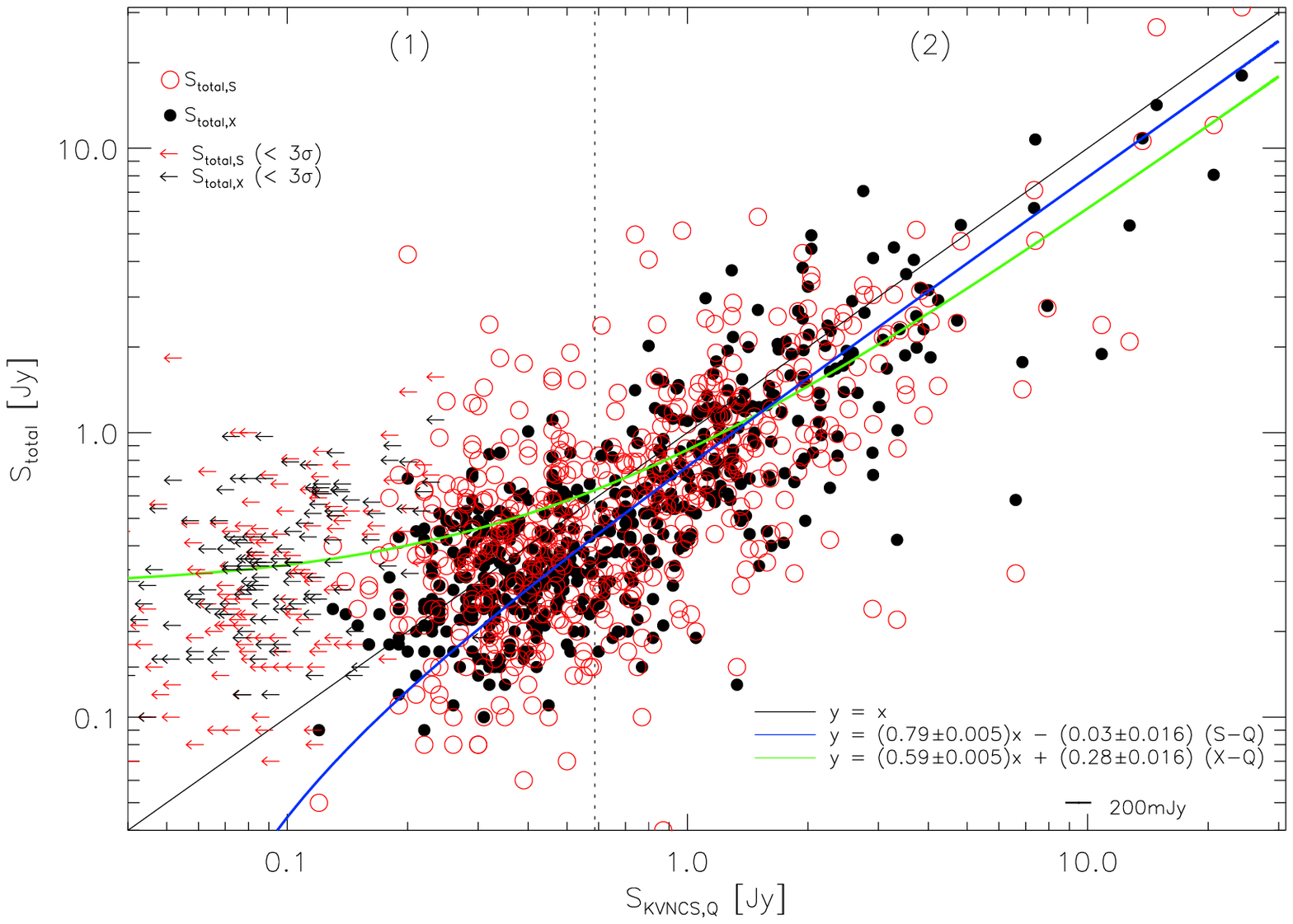}
\caption{Top: the flux--flux relationship between the VCS ($S_{total}$) in the S (red circle) and X (black dot) bands and the measured flux densities of the KVNCS ($S_{KVNCS}$) in the K band. Black and red arrows indicate 1$\sigma$ errors of the sources at less than 3$\sigma$ detection. Black bars on the right and bottom indicate the mean 1$\sigma$ measurement errors (70 mJy) of the bright sources (over 1 Jy) in the K band. Regions (1) and (2) are separated by the median flux densities (the first dotted line) of the KVNCS in the K band. Regions (2) and (3) are separated by 1 Jy (the second dotted line) arbitrarily. The blue (S--K) and green (X--K) lines are linear fit results obtained using the data in (2), and the black line is a proportional case. Bottom: the flux--flux relationship between the VCS ($S_{total}$) in the S (red circle) and X (black dot) bands and the measured flux densities of the KVNCS ($S_{KVNCS}$) in the Q band. Black and red arrows indicate 1$\sigma$ errors of the sources at less than 3$\sigma$ detection. The black bars on the right and bottom indicate the mean 1$\sigma$ measurement errors (200 mJy) of the bright sources over 1 Jy in the Q band. Region (1) and (2) are separated by the median flux densities (the dotted line) of the KVNCS in the Q band. The blue (S--Q) and green (X--Q) lines are linear fit results obtained using the data in (2), and the black line is a proportional case.}
\end{figure} 

\begin{figure} 
\centering
\plotone{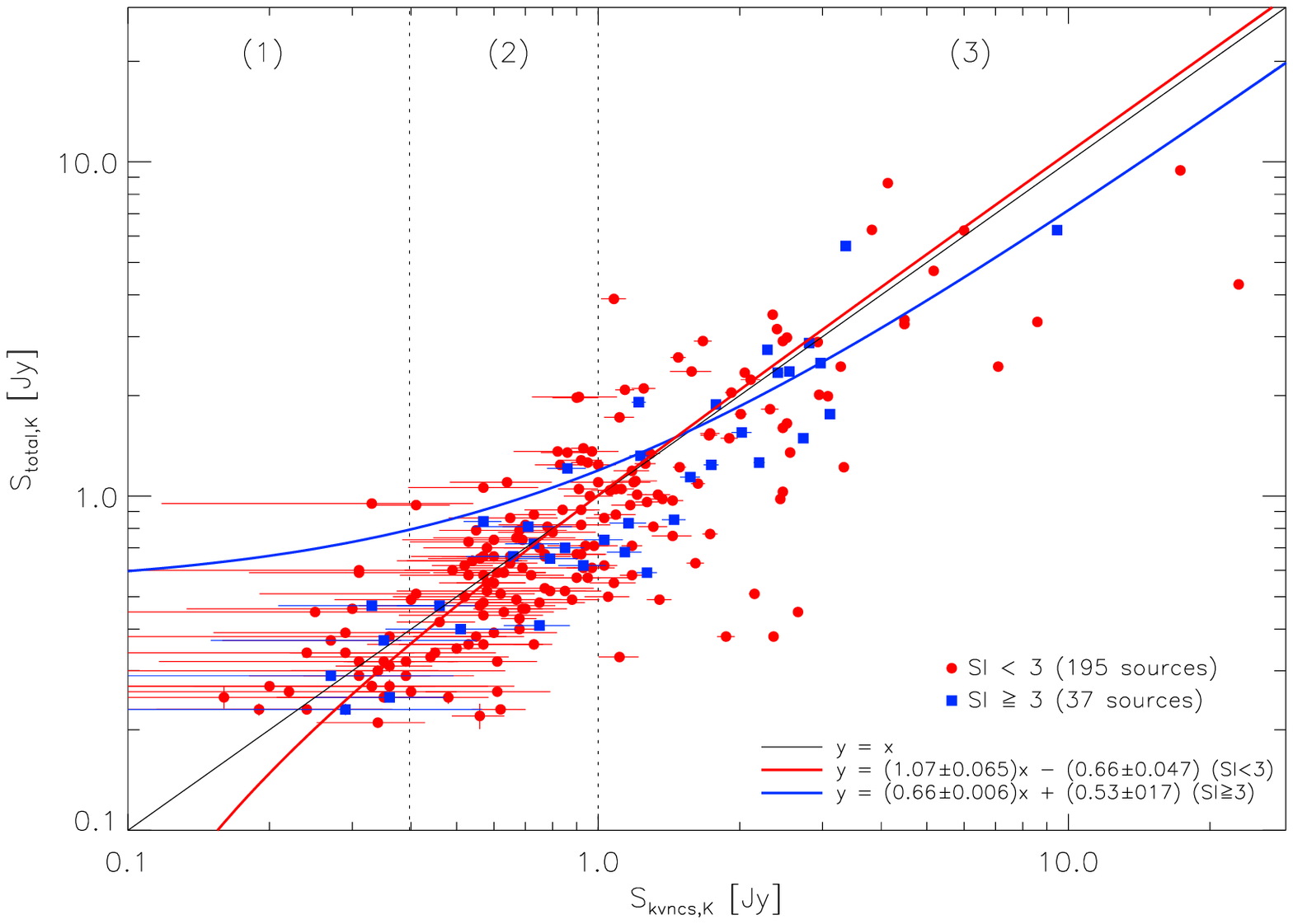}\\
\plotone{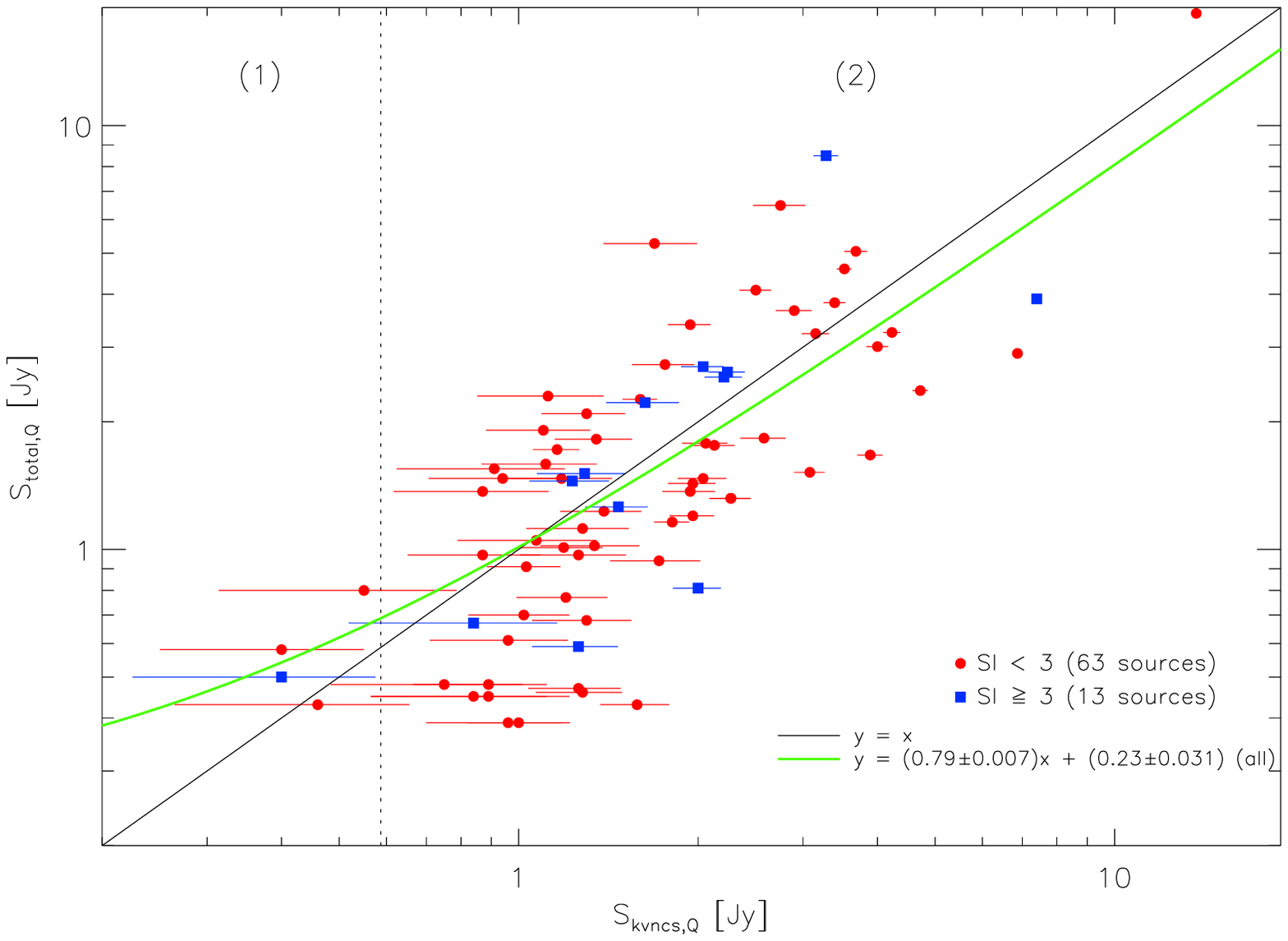}\\
\caption{Top: flux--flux relationship of 232 common sources in the K band between single dish observations from the KVNCS and VLBI observations from \citep{Charlot:2010AJ} according to the structure index (SI). The error bar of each source indicates the 1$\sigma$ measurement error. Regions (1) and (2) are separated by the median flux densities (the first dotted line) of the KVNCS in the K band. Regions (2) and (3) are separated by 1 Jy (the second dotted line) arbitrarily. Red (SI $<$ 3) and blue (SI $\geq$ 3) lines show the weighted linear fit results obtained using the data in (2) and those from all extended sources, respectively (black line: proportional case). Bottom: flux--flux relationship of 76 common sources in the Q band. The error bar of each source indicates the 1$\sigma$ measurement error. Green line shows the weighted linear fit result obtained using the all data, regardless of the SI (black line: proportional case).}
\end{figure}

\begin{figure}
\centering
\plotone{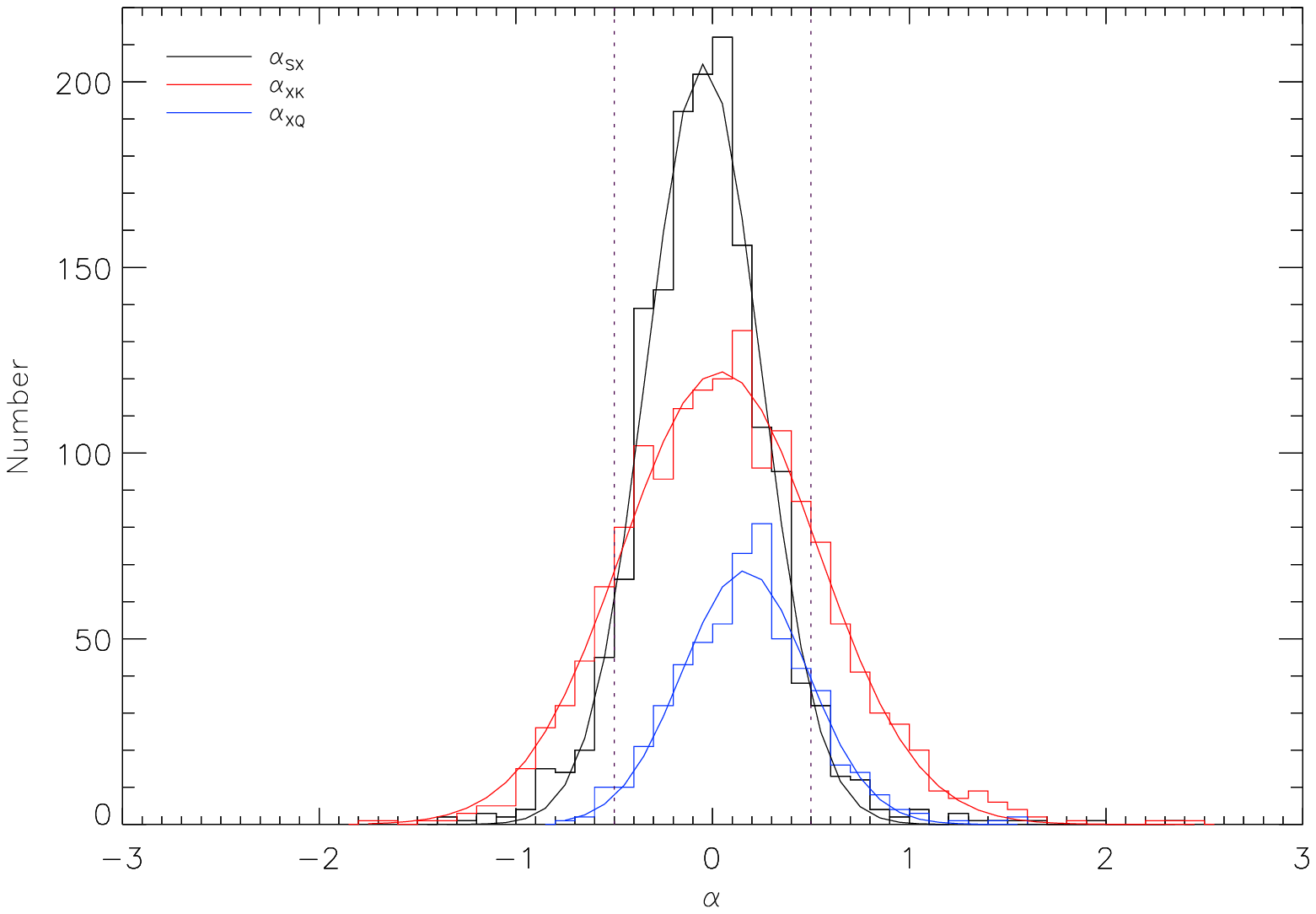}
\plotone{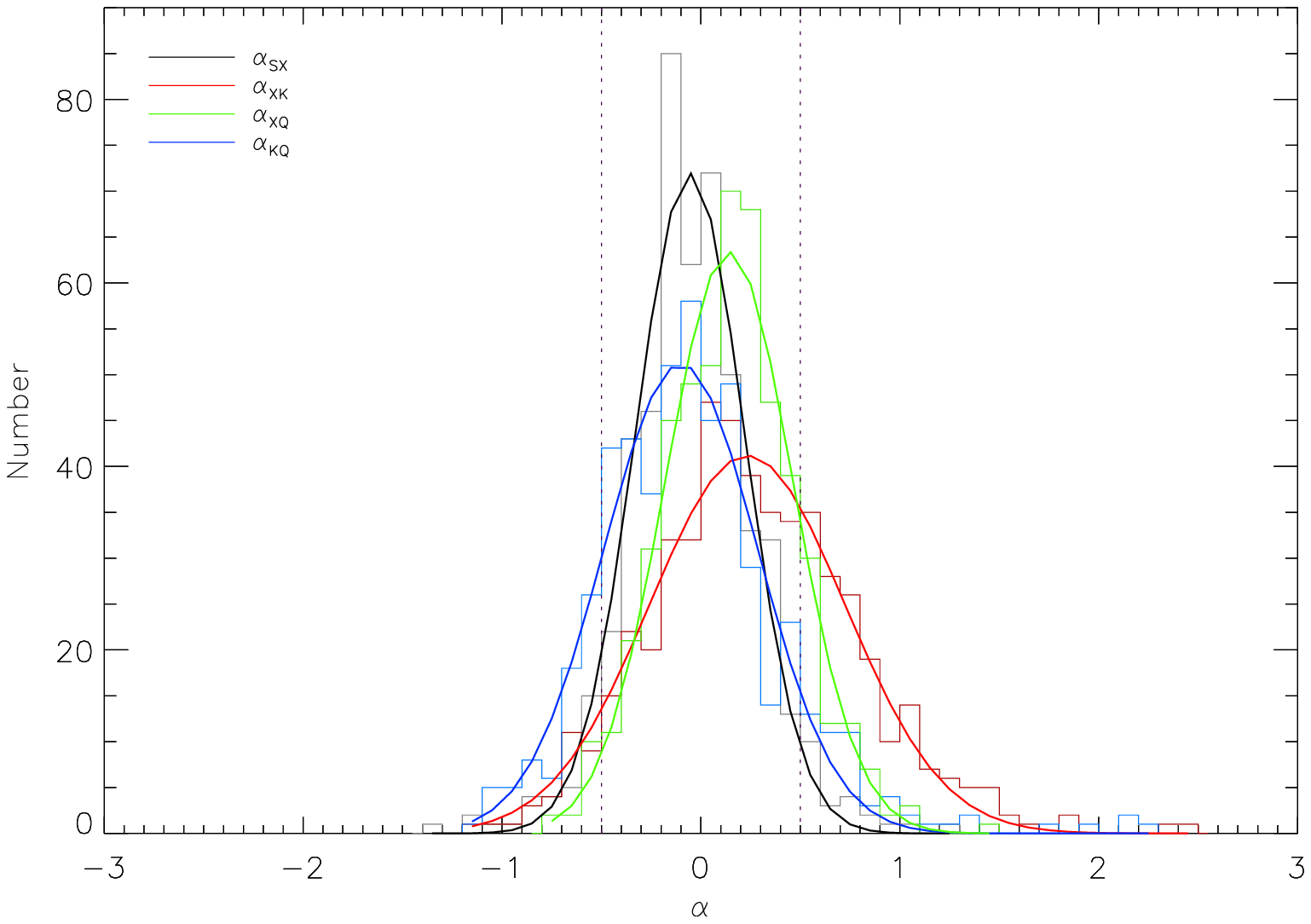}
\caption{Top: number distributions of $\alpha_{SX}$ (black) and $\alpha_{XK}$ (red) of 1533 sources and $\alpha_{XQ}$ (blue) of 553 sources; black, red, and blue lines are Gaussian fit results; coefficients = (peak, center spectral index, standard deviation): $\alpha_{SX}$ = (205, $-$0.045, 0.290), $\alpha_{XK}$  = (122, 0.039, 0.499), and $\alpha_{XQ}$ = (68, 0.165, 0.318). Bottom: number distributions of $\alpha_{SX}$ (black), $\alpha_{XK}$ (red), $\alpha_{XQ}$ (green), and $\alpha_{KQ}$ (blue) of 513 sources measured simultaneously in the K and Q bands; black, red, green, and blue lines are Gaussian fit results; coefficient = (peak, center spectral index, standard deviation): $\alpha_{SX}$ = (72, $-$0.055, 0.275), $\alpha_{XK}$ = (41, 0.23, 0.490),  $\alpha_{XQ}$ = (63.4, 0.14, 0.321), and $\alpha_{KQ}$ = (51, -0.101, 0.387). Vertical purple dotted lines indicate the typical minimum ($\alpha$=-0.5) and maximum ($\alpha$=0.5) $\alpha$ of flat spectrum. VLBI total flux densities in the S and X bands are taken from the VLBA Calibrator Survey (VCS). }
\end{figure}

\begin{figure}
\plotone{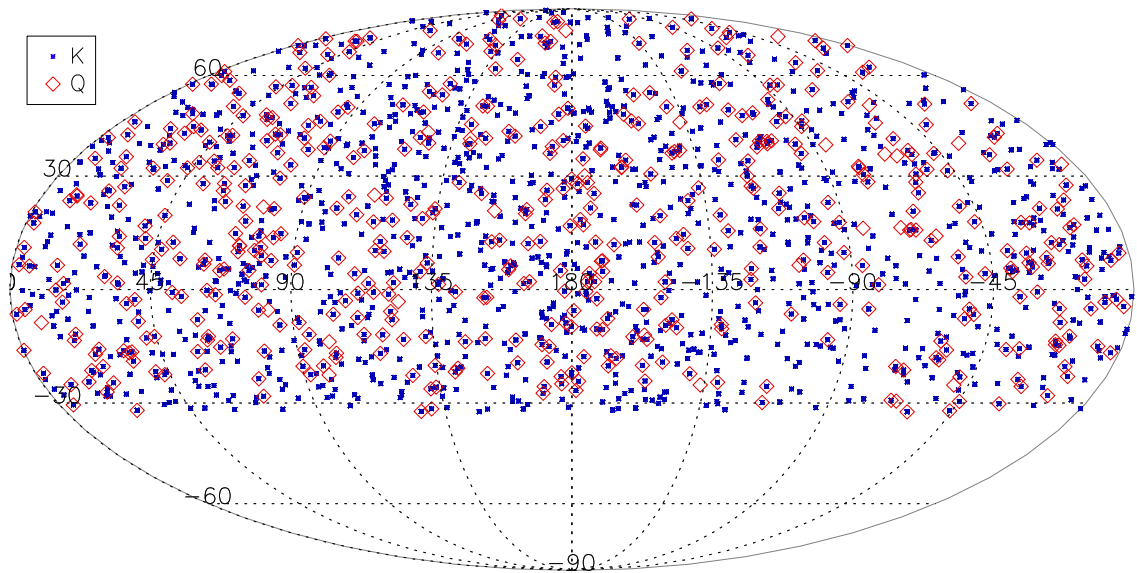}\\
\caption{ Sky distributions of our samples on the celestial sphere: 1533 (K band, blue asterisks) and 553 (Q-band, red diamonds) sources (same as Figure 1 in \citep{jalee:2012PoS}). The lower limit in declination is $-$32.5$\degr$. (Abscissa: right ascension [$\degr$], ordinate: declination [$\degr$]). }
\end{figure}

\begin{figure}
\centering
\plotone{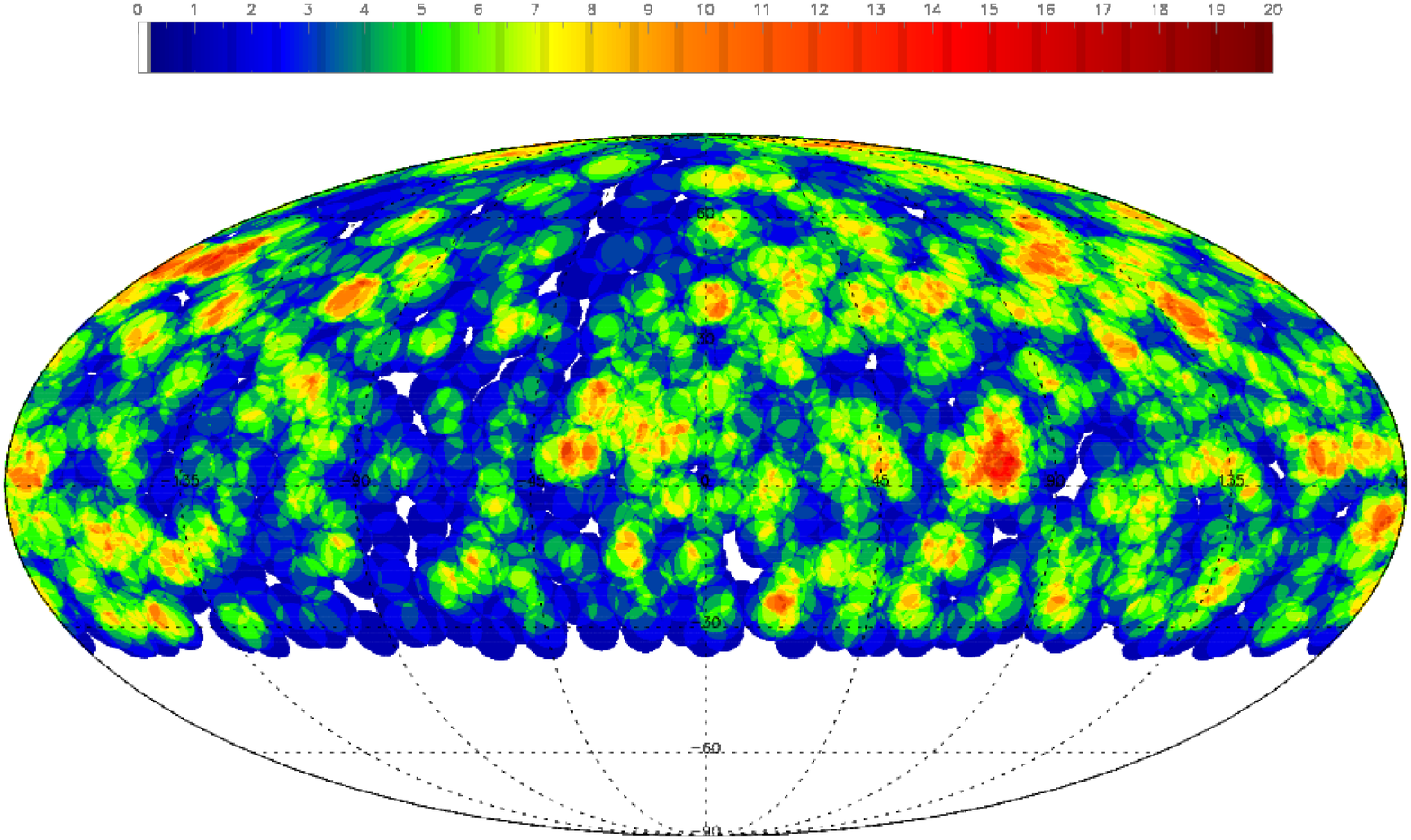}\\
\caption{Sky coverage of 1533 sources by assuming a spatial coherence of 5$\degr$ (radius of circles in Figure 1). Colored legend shows the values represented by overlapping circles. More than two sources are overlapped (89$\%$); a single source (10$\%$); no source (1$\%$) is indicated by white regions. Mollwide equal-area projection technique was used (abscissa: right ascension [$\degr$], ordinate: declination [$\degr$])}
\end{figure}

\clearpage

\begin{deluxetable}{cccccccc}
\tabletypesize{\scriptsize}
\tablecaption{Observational information}
\tablewidth{0pt}
\tablehead{
\colhead{Date} & \colhead{Observation} & \colhead{Number of } & \colhead{Telescope\tablenotemark{$*$}}& \colhead{On-source} & \colhead{Tsys\tablenotemark{$\dagger$}} & \colhead{$\tau_{0}$\tablenotemark{$\dagger\dagger$}} & \colhead{Conversion}\\
& code & sources [$\#$] & & integration time [s] & [K] & & factor [Jy/K] }
\startdata
Dec. 2009 & KVNCS1.0.1 & \nodata & KYS & 40(K), 20(Q) & 96.2(K), 181.7(Q) & 0.053(K), 0.097(Q) & 12.95(K), 12.99(Q)\\
Feb. 2010 & KVNCS1.1.1 & \nodata & KYS  & 40(K), 20(Q) & 74.3(K), 157.4(Q) & 0.027(K), 0.088(Q) & 12.78(K), 11.82(Q) \\
Mar. 2010 & KVNCS1.1.2 & \nodata & KYS  & 40(K), 20(Q) & 97.0(K), 168.9(Q) & 0.068(K), 0.104(Q) & 13.11(K), 12.88(Q) \\
May 2010 & KVNCS1.1.3 & 595 \tablenotemark{$\S$} & KYS  & 40(K), 20(Q) & 220.4(K), 397.1(Q) & 0.103(K), 0.111(Q) & 12.97(K), 12.45(Q) \\
June 2010 & KVNCS1.2.1 & 396 &  KYS  & 80(K), 40(Q) & 124.7(K), 184.0(Q) & 0.111(K), 0.116(Q) & 12.97(K), 12.45(Q) \\
Dec. 2010 & KVNCS1.2.2 & 670 & KUS& 80(K), 40(Q) & 95.8(K), 119.4(Q) & 0.051(K), 0.086(Q) & 13.67(K), 15.33(Q) \\
Jan. 2011 & KVNCS1.3.1 & 573 & KUS & 160(K), 80(Q) & 82.9(K), 137.5(Q) & 0.034(K), 0.092(Q) & 13.85(K), 16.24(Q) \\
Mar. 2011 & KVNCS1.3.2 & 182 & KYS & 160(K), 80(Q) & 88.3(K), 180.2(Q) & 0.042(K), 0.091(Q) & 12.99(K), 13.51(Q) \\
\enddata 
\tablenotetext{*} {KYS: KVN Yonsei Radio Observatory, KUS: KVN Ulsan Radio Observatory}
\tablenotetext{\dagger} {Mean system temperature of each observation}
\tablenotetext{\dagger\dagger} {Mean zenith optical depth of each observation} 
\tablenotetext{\S}{Total number of sources from KVNCS1.0.1 to KVNCS1.1.3 is 595.} 
\end{deluxetable}
\clearpage

\begin{deluxetable}{ccccccccc}
\tablecolumns{9}
\tabletypesize{\scriptsize}
\tablewidth{0pt}
\tablecaption{Examples from the KVNCS Catalog}
\tablehead{
\colhead{IAU name} & \colhead{IVS name} & \colhead{R.A.} & \colhead{Dec.} & \colhead{Obs. code} & \colhead{Flux density ($\pm$ 1$\sigma$)} & \colhead{Obs. code} & \colhead{Flux density ($\pm$ 1$\sigma$)} & \colhead{Redshift (z)} \\
\colhead{(J2000)} & \colhead{}& \colhead{[hh:mm:ss]} & \colhead{[$\degr$:$\arcmin$:$\arcsec$]} & \colhead{ } & \colhead{[Jy]} & \colhead{ } & \colhead{[Jy] } & \colhead{}
}
\colnumbers
\startdata
J0007+5706 &  0005+568 & 00:07:48.47 &  57:06:10.44 &  KVNCS1.3.1 &  0.34 ($\pm$  0.04)  &  KVNCS1.3.1 &  0.23 ($\pm$  0.05) & \nodata \\
J0008-2339 &  0005-239 & 00:08:00.37 & -23:39:18.00 &  KVNCS1.3.1 &  0.16 ($\pm$  0.03)  &   \nodata &   \nodata & 1.412 \\
J0008+6837 &  0005+683 & 00:08:33.47 &  68:37:22.08 &  KVNCS1.3.1 &  0.21 ($\pm$  0.02)  &   \nodata &   \nodata & \nodata \\
J0009+0628 &  0006+061 & 00:09:03.93 &  06:28:21.25 &  KVNCS1.3.1 &  0.22 ($\pm$  0.03)  &  KVNCS1.3.1 &  0.33 ($\pm$  0.05) & \nodata \\
J0009+4001 &  0006+397 & 00:09:04.17 &  40:01:46.56 &  KVNCS1.3.2 &  0.29 ($\pm$  0.04)  &   \nodata &   \nodata & 1.830 \\
J0010+1058 &    IIIZW2 & 00:10:31.00 &  10:58:29.64 &  KVNCS1.0.1 &  2.15 ($\pm$  0.12)  &  KVNCS1.0.1 &  2.90 ($\pm$  0.58) & 0.089 \\
J0010+1724 &  0007+171 & 00:10:33.99 &  17:24:18.72 &  KVNCS1.3.1 &  0.50 ($\pm$  0.04)  &  KVNCS1.3.1 &  0.46 ($\pm$  0.07) & 1.601 \\
J0010-2157 &  0008-222 & 00:10:53.65 & -21:57:04.32 &  KVNCS1.3.1 &  0.41 ($\pm$  0.03)  &  KVNCS1.3.1 &  0.45 ($\pm$  0.07) & \nodata \\
\enddata
\tablecomments{(1) IAU name (J2000), (2) IVS names, (3) and (4) are the right ascension and declination taken from the VCS1$-$5, respectively, (5)  and (6) are an observation code and the source flux density with 1$\sigma$ error in the blanket in the K band, respectively, (7) and (8) are the same as (5) and (6) only in the Q band, (9) redshifts ($z$) of 1137 sources come from NASA/IPAC Extragalactic Database (NED) and Sloan Digital Sky Survey (SDSS) DR13. Full KVNCS catalog is given in the appendix.}
\end{deluxetable}
\clearpage

\begin{deluxetable}{cccccccc}
\tablecolumns{8}
\tablecaption{Statistics of $\alpha$ for 1533 sources in the K band, 553 sources in the Q band, and 513 sources measured simultaneously in the K and Q bands}
\tablewidth{0pt}
\tablehead{
\colhead{} & \multicolumn{3}{c}{1533 (K) $\&$ 553 (Q) sources} & \colhead{} & \multicolumn{3}{c}{513 sources} \\
\cline{2-4} \cline{6-8}
\colhead{Spectral index}& \colhead{Steep} & \colhead{Flat} & \colhead{Invert} &\colhead{} &  \colhead{Steep} & \colhead{Flat} & \colhead{Invert}  \\
\colhead{} & \colhead{[$\%$]} & \colhead{[$\%$]} & \colhead{[$\%$]} & \colhead{} & \colhead{[$\%$]} & \colhead{[$\%$]} & \colhead{[$\%$]} } 
\startdata
$\alpha_{SX}$ &6.9 & 88.5 & 4.6 & &  6.4 & 89.5 & 4.1\\
$\alpha_{XK}$ & 12.9  & 68.2  & 19.9 &&  5.8 & 62.6 & 31.6\\
$\alpha_{XQ}$ & 2.3 & 82.3 & 15.4 && 2.7 & 84.2 & 13.0 \\
$\alpha_{KQ}$ &\nodata  &\nodata  & \nodata && 13.5 & 76.2 & 10.3 \\
\enddata 
\end{deluxetable}
\clearpage

\appendix

\begin{deluxetable}{ccccccccc}
\tablecolumns{9}
\tablenum{2}
\tabletypesize{\scriptsize}
\tablewidth{0pt}
\tablecaption{The KVNCS catalog}
\tablehead{
\colhead{IAU name} & \colhead{IVS name} & \colhead{R.A.} & \colhead{Dec.} & \colhead{Obs. code} & \colhead{Flux density ($\pm$ 1$\sigma$)} & \colhead{Obs. code} & \colhead{Flux density ($\pm$ 1$\sigma$)} & \colhead{Redshift (z)} \\
\colhead{(J2000)} & \colhead{}& \colhead{[hh:mm:ss]} & \colhead{[$\degr$:$\arcmin$:$\arcsec$]} & \colhead{ } & \colhead{[Jy]} & \colhead{ } & \colhead{[Jy] } & \colhead{}
}
\colnumbers
\startdata
J0007+5706 &  0005+568 & 00:07:48.47 &  57:06:10.44 &  KVNCS1.3.1 &  0.34 ($\pm$  0.04)  &  KVNCS1.3.1 &  0.23 ($\pm$  0.05) & \nodata \\
J0008-2339 &  0005-239 & 00:08:00.37 & -23:39:18.00 &  KVNCS1.3.1 &  0.16 ($\pm$  0.03)  &   \nodata &   \nodata & 1.412 \\
J0008+6837 &  0005+683 & 00:08:33.47 &  68:37:22.08 &  KVNCS1.3.1 &  0.21 ($\pm$  0.02)  &   \nodata &   \nodata & \nodata \\
J0009+0628 &  0006+061 & 00:09:03.93 &  06:28:21.25 &  KVNCS1.3.1 &  0.22 ($\pm$  0.03)  &  KVNCS1.3.1 &  0.33 ($\pm$  0.05) & \nodata \\
J0009+4001 &  0006+397 & 00:09:04.17 &  40:01:46.56 &  KVNCS1.3.2 &  0.29 ($\pm$  0.04)  &   \nodata &   \nodata & 1.830 \\
J0010+1058 &    IIIZW2 & 00:10:31.00 &  10:58:29.64 &  KVNCS1.0.1 &  2.15 ($\pm$  0.12)  &  KVNCS1.0.1 &  2.90 ($\pm$  0.58) & 0.089 \\
J0010+1724 &  0007+171 & 00:10:33.99 &  17:24:18.72 &  KVNCS1.3.1 &  0.50 ($\pm$  0.04)  &  KVNCS1.3.1 &  0.46 ($\pm$  0.07) & 1.601 \\
J0010-2157 &  0008-222 & 00:10:53.65 & -21:57:04.32 &  KVNCS1.3.1 &  0.41 ($\pm$  0.03)  &  KVNCS1.3.1 &  0.45 ($\pm$  0.07) & \nodata \\
\enddata
\tablecomments{(1) IAU name (J2000), (2) IVS names, (3) and (4) are R.A. and Declination taken from the VCS1$-$5, respectively, (5) and (6) the observation name and source flux density with 1$\sigma$ errors in the blanket at K band, respectively, (7) and (8) are the same with (5) and (6) at Q band. (9) redshifts (z) of 1137 sources, come from NASA/IPAC Extragalactic Database (NED) and Sloan Digital Sky Survey DR13.}
\tablenotetext{*} {These redshift are taken from SDSS DR13.}
\tablenotetext{\dagger} {The z$-$flag of these sources in NED are PHOT which means the photometric redshift.}
\tablenotetext{\dagger\dagger} {The z$-$flag of these sources in NED are UNCERTAIN which means the uncertain redshift.}

\end{deluxetable}

\end{document}